\documentstyle[12pt,floats,aps]{revtex}

\newcommand{\D}{\displaystyle}
\def\emline#1#2#3#4#5#6{%
       \put(#1,#2){\special{em:moveto}}
       \put(#4,#5){\special{em:lineto}}}
\def\newpic#1{}
\begin{document}
\nonstopmode
\title{On the possibility of fusion reactions in water molecules%
\footnote{ {\rm  LANL E-print {\tt nucl-th/9601021}; 
 Published in}\,\,
{\sl JINR Rapid Communications}, 1995, Nr.~6~[74]--95, p.~5--8}}
\author{V.B. Belyaev, A.K. Motovilov}
\address{Joint Institute for Nuclear Research, Dubna, 141980, Russia}

\author{W. Sandhas}
\address{Physikalisches Institut, Universit\"{a}t Bonn, D-53115 Bonn,
Germany}

\preprint{NUCL-TH/9601021}

\maketitle
\bigskip

\bigskip

\begin{abstract}
\noindent
The probability of nuclear transitions $p+p+{}^{16}{O} \, \rightarrow \,
{}^{18}{Ne}\, (4.522,1^-)$ in molecular water is estimated. Due to the
practically exact agreement of the energy of the $Ne$ resonance and of
the $p+p+{}^{16}{O}$ threshold, the transition probability is found to be
considerably enhanced. This indicates the possibility of nuclear fusion
in rotationally excited $H_2O$ molecules of angular momentum $1^-$.
\end{abstract}
\bigskip

\bigskip

Nuclear states with binding or resonance energies close to breakup
thresholds have a large spatial extension due to the long tail of
the corresponding wave
functions. For instance, the ground state of the nucleus
${}^{8}B$, i.e., of the main source of high-energy solar
neutrinos~\cite{ref1}, is separated from the $p+ {}^{7}{Be}$
threshold by only 130~keV. Integrations
up to 300~fm, hence, are needed \cite{ref2} when treating the
process $p+ {}^7{B}e \longrightarrow {}^8{B}+\gamma$.

In nuclear reactions, the existence of near-threshold intermediate
states leads to a considerable increase of the transition probability.
As an example we recall the muon-catalyzed $dt$ fusion in the molecule
$(dt\mu)$, which takes place primarily via the mechanism
$$
   d+t\longrightarrow {}^5{He}\,(3/2^+)\longrightarrow {}^4{He}+n.
$$
Since the difference between the energies of the $dt$ threshold
and the ${}^5{He}\,(3/2^+)$ resonance is only about
50~keV, it is not surprising that the probability
of this process exceeds at least by four orders of magnitude the
probability of nuclear transitions in the $(dd\mu)$
or $(pd\mu)$ molecules where no such resonances occur \cite{ref3}.

\begin{figure}
\centering
\unitlength=1.10mm
\special{em:linewidth .75pt}
\linethickness{.75pt}
\begin{picture}(77.34,134.33)
\emline{11.00}{81.33}{1}{47.33}{81.33}{2}
\emline{11.00}{120.33}{3}{47.33}{120.33}{4}
\emline{47.33}{120.33}{5}{50.67}{120.33}{6}
\emline{11.33}{96.33}{7}{47.33}{96.33}{8}
\emline{28.00}{81.33}{9}{5.67}{53.67}{10}
\emline{5.61}{53.64}{11}{6.69}{56.21}{12}
\emline{5.57}{53.64}{13}{8.00}{55.18}{14}
\emline{53.96}{120.35}{15}{58.12}{120.35}{16}
\emline{61.14}{120.35}{17}{66.36}{120.35}{18}
\emline{47.33}{109.36}{19}{50.67}{109.36}{20}
\emline{53.96}{109.37}{21}{58.12}{109.37}{22}
\emline{61.14}{109.37}{23}{66.36}{109.37}{24}
\put(29.00,124.33){\makebox(0,0)[cc]{$\Gamma_{p}\leq 20$~keV}}
\put(29.00,116.33){\makebox(0,0)[cc]{4.522$\quad (1^-)$}}
\put(77.34,120.33){\makebox(0,0)[cc]{${}^{16}{O}+2p$}}
\put(61.66,116.33){\makebox(0,0)[cc]{4.522}}
\put(61.33,105.33){\makebox(0,0)[cc]{3.922}}
\put(77.34,109.33){\makebox(0,0)[cc]{${}^{17}{F}+p$}}
\put(37.67,78.67){\makebox(0,0)[cc]{($0^+,\tau=1$)}}
\put(5.00,49.00){\makebox(0,0)[cc]{${}^{18}{F}$}}
\put(29.00,92.34){\makebox(0,0)[cc]{1.8873$\quad (2^+)$}}
\put(19.00,64.34){\makebox(0,0)[cc]{$\beta^+$}}
\emline{11.00}{81.33}{25}{11.00}{134.33}{26}
\emline{47.00}{81.33}{27}{47.00}{134.33}{28}
\end{picture}
\caption{
Fragment of the nucleus ${}^{18}{Ne}$ spectrum.
}
\label{Fig1}
\end{figure}
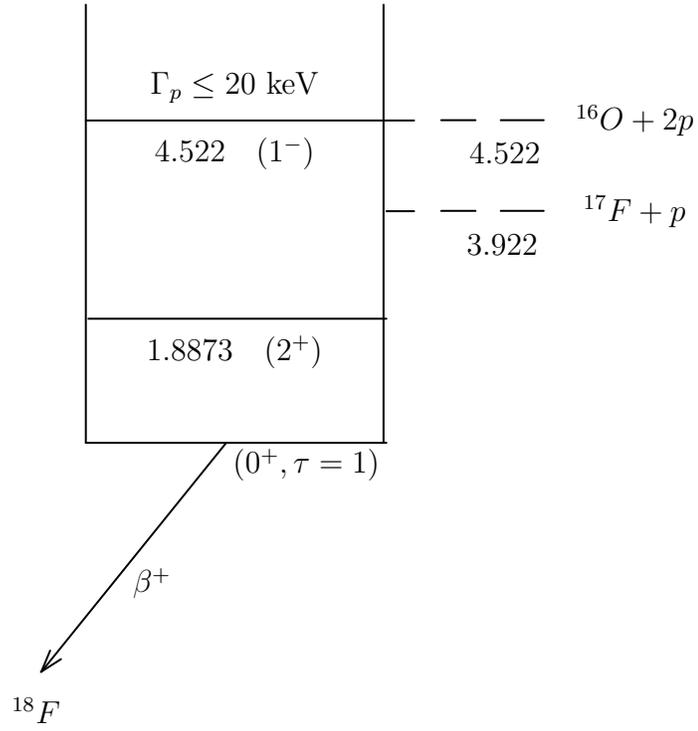

In this note we want to point to an analogous situation, however in an
ordinary (electronic) molecule. From the level scheme of the
nucleus $^{18}{Ne}$  \cite{ref3-p,ref3-pp} presented in
Fig.~\ref{Fig1}, we infer that the
measured energy $E = 4.522 MeV$ of $^{18}{Ne}$ $(1^{-})$
coincides up to the last figure with the threshold energy
of the three-body channel
$p+p+{}^{16}{O}$.  Since the binding energy of the water molecule
is only a few eV, this means that the rotational $1^{-}$ state
of $H_2{O}$ and the $^{18}{Ne}\,(1^-)$ state are degenerate
in energy. Excited molecular water of this angular momentum, thus, is
to be considered as a superposition of these pure molecular and nuclear
states. In other words, the wave function of real water molecules in the
$1^-$ state contains always an admixture of the $^{18}{Ne}$~$(1^-)$
nuclear wave function.

The mixing coefficient of this superposition, and thus the nuclear
transition probability in the $H_2{O}$ molecule, is given by the overlap
integral between the ``pure'' states. Due to the proximity of the
resonance and threshold energies, intermediate and large distances
(in nuclear scale) contribute considerably to this integral.
As a consequence, the nuclear transition probability is enhanced, instead
of being suppressed by the usual Coulomb barrier factor~\cite{ref4}.\\

The estimate on which this statement rests is based on the following
ansatz for the wave function of the water molecule,

\begin{equation}
\label{Eq1}
\psi_{\rm mol}(X)=\D\frac{1}{N_{\rm mol}} \,
\frac{F_{5/2}(\eta_0,\kappa\rho)}{\rho^{5/2}} \,
{\rm e}^{-\kappa\rho} Y^{1M}_{l\lambda}(\hat{x},\hat{y}).
\end{equation}

\noindent
Here we use, instead of the Jacobi variables $\lbrace \vec x, \vec y
\rbrace$ of the $p+p+{}^{16}{O}$ system, the set of hyperspherical
variables $X = \lbrace \rho, \omega, \hat x, \hat y \rbrace$ with
$\rho=\sqrt{x^2+y^2}$ being the hyperradius and \mbox{$\omega = \mbox{arc
tan} \, y / x$} the hyperangle. For the five angles in $X$,
the notation $\Omega = \lbrace \omega,  \hat x, \hat y \rbrace$ will be
used, and the Coulombic potential of our problem is written
in the form $V(X) = {\cal V}(\Omega)/ \rho$. By $F_\nu$ the regular
solutions of the hyperradial Schr\"odinger equation are denoted, and
by $Y^{JM}_{l\lambda}(\hat{x},\hat{y})$ the eigenfunctions of
the total angular momentum operator. $N_{\rm mol}$ is a normalization
factor, and $\kappa\sim\sqrt{|\varepsilon_{\rm mol}|}$ represents the
momentum corresponding to the binding energy $\varepsilon_{\rm mol}$
of the $H_2{O}$ molecule; $\eta_0 = {\cal V}_0 / 2 \kappa$ is a
kind of Sommerfeld parameter, where ${{\cal V}_0}$ is obtained by
averaging ${\cal V}(\Omega)$ with the angular part of
$\psi_{\rm mol}(X)$. The ansatz~(\ref{Eq1}) takes correctly into
account the Coulomb repulsion between the particles at small
distances, as well as the geometric size of the water molecule.

For the description of the ${}^{18}{Ne}$ nuclear resonance state
$(1^-)$, we use the asymptotic form of the Coulombic three-body
break-up function normalized to the nuclear volume,

\begin{equation}
\label{Eq2}
\psi_{\rm res}(X)=\,\D\frac{1}{N_{\rm res}}\,
\frac{f^1(\rho,\omega)}{\rho^{5/2}} \,
Y^{1M}_{l\lambda}(\hat{x},\hat{y}),
\end{equation}
where
\begin{equation}
\label{Eq3}
f^{J}(\rho,\omega)=\int d\hat{x} d\hat{y}
\exp\left\{iK\rho-i\,\D\frac{{\cal V}(\Omega)}{2K}\,\ln(2K\rho) \right\}
Y^{{J}M}_{l\lambda}(\hat{x},\hat{y}).
\end{equation}
Here, $K\sim\sqrt{E}$ is the momentum corresponding to the energy $E$
of the outgoing particles ${}^{16}{O}+p+p$.

Within the models ~(\ref{Eq1}) and~(\ref{Eq2})
we find for the overlap integral the
asymptotic estimate

\begin{equation}
\label{Eq4}
I\sim
\exp\left\{-\D\frac{\pi}{2}\,\eta^0_K\right\}
\exp\left\{ i\eta^0_K S \right\},
\end{equation}
where $\eta_K^0 = {\cal V}_{min} / 2K$ is another kind of Sommerfeld
parameter with ${\cal V}_{min}$ being the minimal value of the
angular part ${\cal V}(\Omega)$ of the total Coulomb potential.
The phase $S$ depends on ${\cal V}_{min}$ and a parameter
$\xi = K / \kappa$. From its definition follows that
$\xi$ can vary between
$0\leq\xi\leq\D\sqrt{\D \Gamma_{p} / |\varepsilon_{\rm mol}|}$
with $\Gamma_{p}$ being the width of the ${}^{18}Ne\,(1^-)$ level
for the decay into the $p+p+{}^{16}{O}$ channel. When studying $S$
as a function of $\xi$ it turns out that there exists
a wide subdomain of values of $\xi$ in which $\mathop{\rm Im}S<0$ and
$|\mathop{\rm Im}S|>\D\frac{\pi}{2}$. That is, the overlap
integral~(\ref{Eq4}), and thus the transition rate

\begin{equation}
\label{Eq5}
W=\kappa c|I|^2
\end{equation}
of the process $H_2{O}\longrightarrow {}^{18}{Ne}\,(1^-)$,
increase exponentially with decreasing $K$ (at small energies
$E\sim K^2$). This is to be contrasted with the usual opposite behavior
of transitions into short-ranged (non-resonant) nuclear states.

The above estimates imply that molecular water in the $1^-$ state
has a non-vanishing probability to go over into the excited state
${}^{18}{Ne}(1^-)$, which then will decay either into the channel
${}^{17}{F}+p+Q_1$ ($Q_1\cong 0.6$~MeV) or into the chain
${}^{18}{Ne}\,(1^-)\longrightarrow {}^{18}{Ne}
+\gamma+Q_2\longrightarrow {}^{18}{F}+e^{+}+\nu$
($Q_2\cong 4.522$~MeV).
Unfortunately, the partial widths of these two transitions
are unknown and, therefore, it is impossible by now to estimate
the whole energy release in the considered process of ``burning''
molecular water.

\medskip

This work was partly supported by the Scientific Division of 
NATO, grant No. 930102.  Two of the authors (V.B.B. and A.K.M.) 
are grateful for financial support to the International Science 
Foundation and Russian Government, grant No.~RFB300.

\end{document}